\documentclass[a4paper,12pt]{article}

\usepackage[T2A]{fontenc}
\usepackage[english]{babel}
\usepackage{amsmath}
\usepackage{amssymb}
\usepackage{cite}
\usepackage[unicode,linktocpage=true,plainpages=false,pdfpagelabels=false]{hyperref}

\newcommand{\dd}{\partial}
\newcommand{\de}{\delta}
\newcommand{\m}{\mu}
\newcommand{\n}{\nu}
\newcommand{\ls}{\left(}
\newcommand{\rs}{\right)}
\newcommand{\al}{\alpha}
\newcommand{\ff}{\varphi}
\newcommand{\be}{\beta}
\newcommand{\ga}{\gamma}
\newcommand{\ta}{\tau}
\newcommand{\La}{\Lambda}
\newcommand{\la}{\lambda}
\newcommand{\ka}{\varkappa}
\newcommand{\ep}{\varepsilon}
\newcommand{\ps}{\psi}

\newcommand{\disn}[2]{$$\displaylines{\refstepcounter{equation}%
            \label{#1}\hskip 1em minus 1em #2\hfilneg}$$}
\newcommand{\nom}{\hfil\hskip 1em minus 1em (\theequation)}
\newcommand{\no}{\hfil \hskip 1em minus 1em\phantom{(\theequation)}%
            \hfilneg\cr\hfilneg\hskip 1em minus 1em\hfil}

\begin{document}

\title{Dark matter as a gravitational effect\\ in the embedding theory approach}

\author{
S.~A.~Paston\thanks{E-mail: pastonsergey@gmail.com}\\
{\it Saint Petersburg State University, Saint Petersburg, Russia}
}
\date{\vskip 15mm}
\maketitle

\begin{abstract}
We discuss the possibility of explaining observations usually related to the existence of dark matter by passing from the general relativity (GR) theory to a modified theory of gravity, the embedding theory proposed by Regge and Teitelboim. In this approach, it is assumed that our space-time is a four-dimensional surface in a ten-dimensional flat ambient space. This clear geometric interpretation of a change of a variable in the GR action leading to a new theory distinguishes this approach from the known alternatives: mimetic gravity and other variants. After the passage to the modified theory of gravity, additional solutions that can be interpreted as GR solutions with additional fictitious matter appear besides the solutions corresponding to GR. Just in that fictitious matter, one can try to see dark matter, with no need to assume the existence of dark matter as a fundamental object; its role is played by the degrees of freedom of modified gravity. In the embedding theory, the number of degrees of freedom of fictitious matter is sufficiently large, and hence an explanation of all observations without complicating the theory any further can be attempted.
\end{abstract}

\newpage

\section{Introduction}
The mystery of the nature of dark matter (DM) is one of the most intriguing problems in modern fundamental physics. The hypothesis of its existence allows simultaneously explaining many contradictions appearing in interpreting observations, from the scale of galaxies to that of the Universe (see, e.g., \cite{1611.09846}).
On a galactic scale, this is the explanation of deviations from the expected motion of stars ("galaxy rotation curves"); on a large scale, this is the explanation of the results of observation of gravitational lensing and baryon acoustic oscillations. Finally, on a cosmological scale, this is the solution of the problem of structure formation and (along with dark energy) the role played in solving the problem of the total mass of the Universe compared with the value corresponding to the critical density.

On the whole, all observation data existing at present (among which the observation of CMB anisotropy plays
the
most
important
role)
are
described
rather
well
by
the \emph{$\La$-Cold Dark Matter} ($\La$CDM)
model
\cite{gorbrub1}, which is currently the standard cosmology model. In its framework, DM can be regarded as nonrelativistic dust-like matter, which produces the same gravitational field as ordinary matter, and the nongravitational interaction of DM with ordinary
matter
is
either
absent
or
undetectably
weak
(see
\cite{2005.03520}
for
the
main
known
DM
properties
following
from the existing observations).

Numerous cases where the introduction of DM is useful makes its existence very probable, although attempts
to
directly
detect
it
gave
no
results
\cite{1509.08767,1604.00014}.
Perhaps the most popular ideas to describe the DM nature
are
currently
the
assumptions
that
DM
consists
of
weakly
interacting
massive
particles
(WIMPs)
\cite{1703.07364}
or is the so-called \emph{fuzzy} DM
\cite{astro-ph/0003365}.
The
failure
of
direct
detection
attempts
is
explained
by
the
weakness
of
the
coupling between DM and ordinary matter. Models
in
which
DM
has
a
self-action
\cite{1705.02358}
are
also
considered
among others, including attempted solutions to the problem of DM density at galaxy centers, which is too high -- the
so-called
"core-cusp"
problem
\cite{1606.07790}.

However, the fact that it has not been possible to detect DM in any interaction besides the gravitational one suggests that it is can be an effect of the description of gravity rather than real matter, i.e., in fact, DM does not exists (from the standpoint of fundamental theory). Apparently, this was first done within
the
modified
Newtonian
dynamics
(MOND)
paradigm
\cite{mond}.
In the framework of this approach, the above-mentioned contradiction can be eliminated rather successfully on galactic or similar scales; but on a
cosmological
scale,
the
MOND
paradigm
does
not
work
equally
well
\cite{1910.04368}.

Modifications of the theory of gravity that introduce additional degrees of freedom compared to general relativity (GR) seem much more promising. If we discuss solutions of such modified theories from the standpoint of GR, then additional gravitational degrees of freedom describe the dynamics of some fictitious matter that can be identified with DM. We emphasize that in this case, DM has independent dynamics, i.e., it can move in a way that is di.erent from that of ordinary matter. Modified theories of gravity that have additional degrees of freedom compared to GR include, for example, $f(R)$ gravity, scalar-tensor theories of gravity,
and
a
number
of
others
(see
\cite{1108.6266}).
We note that after passing to a modified theory of gravity, DM \emph{no longer exists} as an independent fundamental object, and only when we discuss this theory in terms of GR does DM arises as a source of an additional contribution to the Einstein equations.

Among the modified theories of gravity with additional degrees of freedom, in the context of the description of DM, the \emph{mimetic} theory of gravity \cite{mukhanov}
appears
to
have
been
discussed
most
often
in
the
last decade. The name of the theory reflects the fact that its gravitational degrees of freedom "mimic" the presence of some fictitious matter. In the simplest version of the theory, this matter is dust-like and moves potentially, i.e., in a vortex-free manner. Mimetic gravity is obtained from GR by replacing the independent variable
\disn{v1}{
g_{\m\n}=\tilde g_{\m\n}\tilde g^{\ga\de}(\dd_\ga\la)(\dd_\de\la)
\nom}
in the GR action
\disn{v2}{
S=S^{\text{EH}}+S_{\text{m}},\qquad
S^{\text{EH}}=-\frac{1}{2\ka}\int\! d^4x\sqrt{-g}\,R,
\nom}
where $S^{\text{EH}}$ is the Einstein-Hilbert action (we use the signature $+---$) and $S_{\text{m}}$ is the action of ordinary
matter. In
\eqref{v1},
$\tilde g_{\m\n}$ is the auxiliary metric (and $\tilde g^{\m\n}$ is the metric inverse to it), which is a new independent variable along with the scalar field $\la$. Because the common factor $\tilde g_{\m\n}$, obviously, does not affect $g_{\m\n}$, ten independent field variables exist in the new theory, as originally in GR; this is usually referred to as "isolating the conformal mode" of the metric.

The variation of the action with respect to the new variables $\tilde g_{\m\n}$ and $\la$ gives equations of motion that can be written as
\disn{vi18}{
G^{\m\n}=\ka \ls T^{\m\n}+n u^\m u^\n\rs,
\nom}\vskip -2em
\disn{vi18.2}{
D_\m (n u^\m)=0,
\nom}
where $G^{\m\n}$ is the Einstein tensor, $T^{\m\n}$ is the energy-momentum tensor (EMT) of ordinary matter, and $D_\m$ is the covariant derivative, and the notation
\disn{vi17}{
n\equiv g_{\m\n}\ls \frac{1}{\ka}G^{\m\n}-T^{\m\n}\rs,
\nom}\vskip -2em
\disn{vi17.2}{
u_\m\equiv\dd_\m\la,
\nom}
is used. It can be verified that the relation
\disn{vi18.3}{
g^{\m\n}u_\m u_\n=1.
\nom}
is satisfied identically. If
this
identity
and
notation
\eqref{vi17}
are
taken
into
account,
then
among
the
ten
Einstein
equations
in
\eqref{vi18},
one
(obtained
by
contraction
with
$g_{\m\n}$) is satisfied identically, which corresponds to the invariance of the action under the Weyl transformation of the auxiliary metric $\tilde g_{\m\n}$.

However, the obtained equations can be interpreted somewhat differently by assuming that $n$ is some additional dynamical variable obeying Eq. \eqref{vi18.2}
and
that
relation
\eqref{vi17}
for
it
is
the
tenth
Einstein
equation
(precisely the one that follows by contraction with $g_{\m\n}$). The physical meaning of this variable is easy to
understand if we recall that $n u^\m u^\n$,
which
is
contained
in
the
right-hand
side
of
Einstein
equations
\eqref{vi18}
as
the EMT of some additional matter, is the EMT of dust-like matter with the density $n$ and velocity $u^\m$ (take
that
the
velocity
normalization
condition
\eqref{vi18.3}
into
account).
Thus,
it
turns
out
that
mimetic
gravity
is
completely
equivalent
to
GR
with
additional
fictitious
matter
that
moves
potentially
(this
follows
from
\eqref{vi17.2}).
Interestingly,
Eq.~\eqref{vi18.2},
which
plays
the
role
of
the
equation
of
motion
for
fictitious
matter,
turns
out
to
be
a
corollary
of
Einstein
equation
\eqref{vi18},
as
is
well
known
for
dust-like
matter.

As we can see, as a result of the change of a variable in the GR action, despite the number of independent variables remaining unchanged, new dynamical variables appear in the theory when interpreting it from the GR standpoint, namely, the density of fictitious matter $n$ and the scalar $\la$ parameterizing the potential velocity. The
reason
for
this
is
the
presence
of
the
differentiation
operation
in
\eqref{v1};
as a result of such a \emph{differential change of variables},
the
theory
dynamics
can
be
enriched
\cite{statja60}.
As
regards
attempts
to
explain
the effects associated with DM in the framework of such an approach, it is important to emphasize that, the fictitious matter described as a gravitational effect has its own dynamics (dynamical degrees of freedom) and hence its own initial data. Depending on the initial data, we can have solutions that are completely equivalent to GR (if we have $n=0$ in some region of space at the initial instant, which allows, for example, describing the galaxies where DM is virtually absent) as well as solutions for which the fictitious matter moves quite differently from ordinary matter; this behavior of DM is known from gravitational lensing observations
of
the
Bullet
cluster
\cite{bulletcluster}.

For theories arising as a result of changing a variable in the GR action, there is usually a possibility of reformulating them in the GR form with additional fictitious matter on the level of not only the equations of motion but also the action. For mimetic gravity, an equivalent reformulation can be obtained by considering GR
with
an
additional
contribution
to
the
action
of
the
form
\cite{Golovnev201439}
\disn{p12}{
S^{\text{add}}=-\frac{1}{2}\int\! d^4 x\, \sqrt{-g}\,\Bigl( 1-g^{\m\n}(\dd_\m\la)(\dd_\n\la)\Bigr)n,
\nom}
with total action
\disn{vv2}{
S=S^{\text{EH}}+S_{\text{m}}+S^{\text{add}}.
\nom}
Here, the independent variables are the usual metric $g_{\m\n}$ and two scalar fields $n$ and $\la$ describing fictitious matter, which still play the role of matter density and velocity potential. Most often, mimetic gravity is studied precisely in this formulation. There are other possibilities of choosing the contribution to the action of
a
potentially
moving
ideal
fluid
without
pressure;
various
options
were
discussed
in
\cite{statja48}.

To reproduce the DM properties that are needed to explain the existing observations in the mimetic gravity framework, the theory has to be made somewhat complex. For example, the addition of a scalar field potential $\la$ to the action has been considered, leading to the appearance of nonzero pressure in fictitious matter
\cite{mukhanov2014}.
The possibility of adding a contribution with higher derivatives of this field has also
been
studied,
turning
fictitious
matter
into
a
nonideal
liquid
\cite{mukhanov2014,vikman2015,kobayashi2017}.
There is also a possibility to transform
the
contribution
to
action
\eqref{p12}
such
that
the
fictitious
matter,
while
remaining
dust-like,
moves
nonpotentially
\cite{statja48}.
In the last case, in addition to the fields $n$ and $\la$, another two scalar fields participate in the description of fictitious matter; it is then possible to return to the original formulation of mimetic gravity with the independent auxiliary metric $\tilde g_{\m\n}$ and three scalar fields contained in the corresponding expression for the physical metric $g_{\m\n}$,
which
is
analogous
to
\eqref{v1}
(see
\cite{mimetic-review17}
and
the
references
therein
for
the
current status of the mimetic gravity approach).

The fact that mimetic gravity is insufficiently "complex" (the fictitious matter corresponding to it has too simple dynamics and the theory has to be made more complex to explain the effects associated with DM) can be viewed as a disadvantage of the approach. Another possible disadvantage is the artificial form
of
the
change
of
a
variable
in
\eqref{v1},
which
underlies
the
approach:
it
is
difficult
to
formulate
any
physical
or geometrical arguments in favor of just this type of replacement. Both these disadvantages are absent in another modified theory of gravity, which had appeared earlier and which also arises as a result of changing the variable in the GR action. This change of a variable
 \disn{r1}{
g_{\m\n}=(\dd_\m y^a)(\dd_\n y^b)\,\eta_{ab}
\nom}
has a clear geometrical meaning: the new independent variable $y^a(x^\al)$ is a function of the embedding of the four-dimensional surface into the ambient space with a flat metric $\eta_{ab}$ (the indices $a,b,\ldots$ label the
components
of
the
Lorentz
coordinates
of
the
ambient
space),
and
\eqref{r1}
defines
the
induced
metric
on
the embedded surface. Thus, this modification of gravity is based on a simple assumption: our space-time is a four-dimensional surface in some flat pseudo-Euclidean ambient space. We note that in the GR framework, our space-time is considered an abstract pseudo-Riemannian space. Such a modified theory of gravity
was
proposed
in
1975
\cite{regge}
and
is
called
Regge-Teitelboim
gravity
or
the
embedding
theory or \emph{embedding gravity}.
The idea of the approach was obviously suggested by the geometric description of strings, which was reflected in
the
title
of
original
paper
\cite{regge}:
\emph{"General relativity `a la string: a progress report"}. The difference between the embedding theory and the Nambu-Goto string theory lies in the dimension under consideration (1 + 3 instead
of
1+1)
and
in
the
choice
of
the
action:
for
the
embedding
theory,
GR
action
\eqref{v2}
is
used,
and
for
the
string, the volume (area in a two-dimensional space) of the manifold is used instead of the Einstein-Hilbert action.

As shown below, the embedding theory is sufficiently "complex" to explain the effects associated with DM in the framework of such an approach without further complicating the theory. In
Secs.~2
and~3,
we describe the embedding theory in more detail, including the possibility of formulating it in GR form with a contribution of additional fictitious matter to the action. In
Sec.~4,
we
discuss
the
weak
gravity
limit for the embedding theory and show that to satisfy the superposition principle for the gravitational field, the background embedding corresponding to the flat metric must correspond to the generic case and, in particular, not be reducible to a four-dimensional plane in the ambient space. In
Sec.~5,
we
present
the
results of considering the nonrelativistic limit of the embedding theory, in which fictitious matter behaves like a nonrelativistic fluid with some self-action. Further studies of the properties of this liquid and their comparison with the observed DM properties will allow deciding whether the passage to the description of gravity in the form of the embedding theory can explain, any additional modifications, the observed effects that are currently explained by the hypothesis stating the DM existence.

\section{Embedding theory: Regge-Teitelboim gravity}
When describing gravity in the form of an embedding theory, the embedding function $y^a(x^\al)$ in terms of which $g_{\m\n}(x^\al)$
is
expressed
by
as
an
induced
metric,
Eq.
\eqref{r1},
is
an
independent
variable
instead
of
the
metric $g_{\m\n}(x^\al)$. A parameter characterizing this approach, the ambient space dimension $N$, then appears. In addition, in principle, there is some arbitrariness in the choice of its signature. If we assume that it be possible to define an arbitrary space-time metric in terms of the embedding theory (at least locally, i.e., for some part of space-time), then a restriction on N arises in accordance with the Janet-Cartan-Friedman
theorem
\cite{fridman61}.
This theorem states that an arbitrary Riemannian space of dimension d can be locally isometrically embedded into any Riemannian space of the dimension
 \disn{v3.1}{
N\geqslant\frac{d(d+1)}{2},
\nom}
and hence, in particular, into the flat space of such a dimension. The theorem was generalized to the pseudo-Riemannian
case
by
Friedman
\cite{fridman61};
then
in
addition
to
restriction
\eqref{v3.1}
on
the
total
dimension
$N$, an additional natural constraint arises, that the number of both spatial and temporal directions in the ambient space be not less than that in the embedded pseudo-Riemannian space.

Because $d=4$
for
our
space-time,
restriction
\eqref{v3.1}
gives
$N\geqslant10$, and the embedding theory with the ambient space dimension $N=10$ is considered most often (it is amazing that this is the same dimension in which superstring theory becomes consistent, but the reason for this coincidence is entirely unclear). This
a
value is the most natural because the numbers of the old variables $g_{\m\n}$ and of the new variables $y^a$ the change
of
variables
\eqref{r1}
then
coincide.
The
ambient
space
signature
must
be
such
that
it
have
at
least
one
timelike direction. Just one timelike direction is typically used; in this case, causality can be established in the ambient space: there are no closed timelike lines, which is important for the embedding theory in the form
of
field
theory
in
the
ambient
space-time
\cite{statja25}.
Thus, it turns out to be most natural to choose the ten-dimensional Minkowski space $R^{1,9}$ as the ambient space for the embedding theory.

As
the
action
of
the
embedding
theory,
GR
action
\eqref{v2}
is
taken,
with
the
induced
metric
\eqref{r1}.
Varying
a
with respect to the new independent variable $y^a$ yields
the
Regge-Teitelboim
equations
\cite{regge},
which
can
be
written in two equivalent forms,
\disn{15}{
D_\m\Bigl(
\ls G^{\m\n}-\ka\, T^{\m\n} \rs \dd_\n y^a\Bigr)=0,
\nom}
or
\disn{15a1}{
\ls G^{\m\n}-\ka\, T^{\m\n} \rs b^a_{\m\n}=0,
\nom}
where $b^a_{\m\n}=D_\m\dd_\n y^a$ is the second fundamental form of a surface. We note although the Einstein equations
contained
the
second
derivatives
of
the
metric
and
change
of
variables
\eqref{r1}
contains
differentiation,
Eqs.~\eqref{15a1}
contain no derivatives of $y^a$ an order higher than two. To
see
this,
it
suffices
to
use
the
formula
\cite{statja18}
 \disn{s1-36.1}{
G^{\m\n}=
\frac{1}{2}\,g_{\xi\zeta}E^{\m\xi\al\be}E^{\n\zeta\ga\de}\,
b^a_{\al\ga} b^b_{\be\de}\,\eta_{ab},
\nom}
where $E^{\m\xi\al\be}=\ep^{\m\xi\al\be}/\sqrt{|g|}$ is the covariant unit antisymmetric tensor.

As we can see, the Regge-Teitelboim equations contain more solutions than the Einstein equations: besides the Einstein solutions, there are others. The resulting extension of the theory dynamics is a consequence
of
the
presence
of
differentiations
in
the
change
of
variables
\eqref{r1},
just
as
in
mimetic
gravity
discussed
in
Sec.~1.
As a result, in the embedding theory, as in mimetic gravity, there are more dynamical variables than in GR, and these variables can be assumed to describe some fictitious matter, which can be identified with DM. To isolate these variables, it is convenient to rewrite the Regge-Teitelboim equations in the form of
the
equivalent
set
of
equations
\cite{pavsic85}:
 \disn{s4}{
G^{\m\n}=\ka \ls T^{\m\n}+\ta^{\m\n}\rs,
\nom}\vskip -2em
 \disn{s5}{
D_\m\Bigl(\ta^{\m\n}\dd_\n y^a\Bigr)=0 \quad\Leftrightarrow\quad
\ta^{\m\n}b^a_{\m\n}=0.
\nom}
(16)
The first of them is the Einstein equation with an additional contribution of the EMT $\ta^{\m\n}$ of fictitious matter. The second can be viewed as an equation restricting the possible behavior of $\ta^{\m\n}$, and therefore as an equation of motion of this fictitious matter. Thus, we can assume that fictitious matter is described
a
by the variables $y^a$ and $\ta^{\m\n}$. Not all of these twenty variables are dynamical, i.e., have their own arbitrary initial data. It is difficult to determine the number of dynamical variables corresponding to fictitious matter and somehow isolate them in the general case, but this can be done in the nonrelativistic limit, which is discussed
in
Sec.~4.
It is also possible to calculate the number of the degrees of freedom of fictitious matter by studying the canonical (Hamiltonian) formulation of the theory (see below).

As
was
already
noted
in
Sec.~1,
theories
arising
as
a
result
of
changing
the
variable
in
the
GR
action
can usually be reformulated in the GR form with additional fictitious matter not only at the level of the
equations
of
motion
but
also
at
the
level
of
the
action
if
the
total
action
in
form
\eqref{vv2}
is
written
with
some $S^{\text{add}}$. This can also be done for the embedding theory in different ways, using which can be convenient in analyzing the properties of fictitious matter in different cases. The simplest way is to write $S^{\text{add}}$ in the form \cite{statja48}
\disn{za1}{
S^{\text{add}}=\frac{1}{2}\int\! d^4 x\, \sqrt{-g}\,
\Bigl( (\dd_\m y^a)(\dd_\n y_a) - g_{\m\n}\Bigr)\tau^{\m\n}.
\nom}
With this choice of $S^{\text{add}}$, the fictitious matter EMT $\ta^{\m\n}$ (which we assume to be symmetric) is a Lagrange
a
multiplier, the variation over which yields the relation of the metric $g_{\m\n}$ to the embedding function $y^a$ accordance
with
condition
\eqref{r1}.
An alternative way of writing $S^{\text{add}}$, with a set of conserved currents chosen an independent variable instead of $\ta^{\m\n}$,
is
described
in
Sec.
3.

After
the
appearance
of
\cite{regge},
the
ideas
of
the
embedding
theory
were
critically
discussed
in
\cite{deser};
subsequently,
they
were
repeatedly
used
to
describe
gravity,
including
its
quantization
(see,
e.g., \cite{maia89,estabrook1999,davkar,statja25,faddeev,statja33}).
At first, the embedding theory was mainly regarded as a reformulation of GR that was potentially more convenient for quantization because of the presence of a flat ambient space; therefore, the presence of non-Einstein
solutions
in
it
was
considered
a
disadvantage.
To
eliminate
it,
the
proposal
in
\cite{regge}
was
to
impose
additional constraints making the theory equivalent to GR; the canonical formalism for the resultant theory was then studied. These
studies
were
continued
in
\cite{statja18,statja24,statja35}, and
the
Hamiltonian
description
of
the
complete
embedding
theory
was
investigated
in
\cite{tapia,frtap,statja44,rojas20,statja72}.
Such a description allows, in particular, determining the number of degrees of freedom of the embedding theory: it turns out to be equal to six, i.e., in
comparison
with
GR,
there
are
another
four
degrees
of
freedom
corresponding
to
fictitious
matter
\cite{statja72}
(we mean four pairs of conjugate canonical variables, which in the Lagrangian language correspond to four variables whose time evolution is controlled by a second-order differential equation). With the appearance of the DM problem, interest in the study of the complete embedding theory increased because non-Einstein solutions
can
be
used
to
explain
the
DM
effects
\cite{davids01,statja26,statja76}.
A somewhat outdated but very detailed list of references
related
to
the
embedding
theory
and
similar
problems
can
be
found
in
\cite{tapiaob}.

\section{Alternative form of the action}
We
note
that
in
its
first
form,
Eq.~\eqref{s5},
which
can
be
understood
as
the
equation
of
motion
of
fictitious
matter, expresses the conservation of a set of currents labeled with the subscript $a$,
\disn{nnn1}{
\dd_\m \ls \sqrt{-g}\,j^\m_a\rs=0,
\nom}
where
\disn{nnn2}{
j^\m_a=\ta^{\m\n}\dd_\n y_a.
\nom}
The description of fictitious matter in the language of such a set of currents is useful for a better understanding of its properties. It is therefore interesting to write the action of fictitious matter with $j^\m_a$ considered as
a one of the variables describing it instead of $\ta^{\m\n}$. We note that doing so is not obstructed by the fact that $j^\m_a$ contains more components (40 components) than $\ta^{\m\n}$ does (10 components).

The
corresponding
action
has
the
form
\cite{statja51}
\disn{r5}{
S^{\text{add}}=\int\! d^4 x\, \sqrt{-g}\,
\Bigl( j^\m_a\dd_\m y^a-\text{\bf tr}\sqrt{g_{\m\n}j^\n_a j^{\al a}}\Bigr),
\nom}
where a square root of a matrix with the indices $\m$ and $\al$ is understood, followed by taking the trace
a
(the operation \text{\bf tr}). Fictitious matter is described, in addition to $j^\m_a$, also by the embedding function $y^a$,
a
which becomes a Lagrange multiplier in this approach. Varying with respect to it gives the condition
\disn{r6}{
D_\m j^\m_a=0.
\nom}
The variation with respect to $j^\m_a$ leads to another equation of motion,
\disn{r7}{
\dd_\m y^a=\hat\be_{\m\n}j^{\n a},
\nom}
where the symmetric tensor $\hat\be_{\m\n}$ is inverse to $\be^{\m\n}$, whose components $\be_\m{}^\al$ are defined as the square root of the matrix $g_{\m\n}j^\n_a j^{\al a}$ with the indices $\m$ and $\al$. The square root if defined in terms of its Taylor expansion
a
about
the
unit
matrix
(see
\cite{statja51}
for
more
details).
It
is
easy
to
verify
that
\eqref{r7}
implies
the
induced-metric
condition \eqref{r1}:
\disn{r8}{
(\dd_\m y^a)(\dd_\n y_a)=\hat\be_{\m\al}j^{\al a}  \hat\be_{\n\be}j^{\be}{}_a=
\hat\be_{\m\al}\hat\be_{\n\be}\be^{\al\ga}\be_\ga{}^\be=g_{\m\n}.
\nom}

Because
action
\eqref{r5}
is
a
contribution
to
the
total
action
\eqref{vv2},
variation
\eqref{r5}
with
respect
to
$g_{\m\n}$ gives
the EMT of fictitious matter $\ta^{\m\n}$. It
can
be
shown
\cite{statja51}
that
$\ta^{\m\n}=\be^{\m\n}$ is obtained as a result of the variation. It is then possible to express $j^\m_a$ from equation of motion \eqref{r7},
with the result coinciding with~\eqref{nnn2}.
In
addition,
because
equation
of
motion
\eqref{r6}
coincides
with
Eq.~\eqref{nnn2},
we conclude that GR with
additional
contribution
\eqref{r5}
to
the
action
in
\eqref{vv2}
completely
reproduces
the
equation
of
motion
of
the
embedding theory.

It
is
useful
to
see
how
action
\eqref{r5}
is
simplified
for
the
set
of
currents
$j^\m_a$ where all currents are zero
a except the one corresponding to $a=0$ ($j^\m_a=j^\m\de^a_0$); the same result can be obtained if the dimension of
a
the ambient space is reduced to $N=1$. For a matrix of unit rank (which is the rank of the matrix in the radicand
in
\eqref{r5}),
the
trace
of
its
square
root
coincides
with
square
root
of
the
trace,
whence
we
obtain
a simplified action in the form
\disn{a11.2a}{
\tilde S^{\text{add}}=\int\! d^4 x\, \sqrt{-g}\,
\Bigl( j^\m\dd_\m y^{0}-\sqrt{j^\m g_{\m\n} j^\n}\,\Bigr).
\nom}
This expression is one of the possibilities to write the action of a potentially moving ideal fluid without pressure
\cite{statja48},
which,
as
was
mentioned
in
Sec.~1,
is
the
fictitious
matter
of
mimetic
gravity.
Therefore, it can be said that mimetic gravity is a certain limit of the embedding theory, and the complete embedding theory is more complex. However, if we restrict the class of fields in the action, the set of variations of these fields also decreases; this means that some of the equations of motion are lost. The result of a more accurate
analysis
in
the
nonrelativistic
limit
of
the
embedding
theory
is
described
in
Sec.~5.

\section{The weak gravity limit}
The gravitational field is considered weak if the metric has the form
 \disn{s20}{
g_{\m\n}=\eta_{\m\n}+h_{\m\n},\qquad
h_{\m\n}\ll 1,
\nom}
where $\eta_{\m\n}$ is the Minkowski space metric. When describing gravity in the framework of the embedding a
theory, a question arises: how to choose the background embedding function $\bar y^a$ corresponding to the
a
background metric $\eta_{\m\n}$? It is obvious that the choice of $\bar y^a$ involves arbitrariness, as is evidenced by the well-known fact that in three-dimensional space, a part of a cylinder has the same flat metric as a part of the plane.

The simplest choice of the background embedding function corresponds to the four-dimensio\-nal plane,
 \disn{nnn4}{
\bar y^a(x^\m)=\de^a_\m x^\m.
\nom}
However, if solutions are sought in the form of small deviations from such a background, i.e., as
 \disn{nnn5}{
y^a=\bar y^a+\de y^a,
\nom}
then
equations
of
motion
\eqref{15a1}
of
the
embedding
theory
are
nonlinear
in
the
deviations
$\de y^a$ (they are cubic, as
can
be
seen
from
the
expression
for
the
Einstein
tensor
in
form
\eqref{s1-36.1}).
Thus, the embedding theory is nonlinearized
on
background
\eqref{nnn4},
which
was
already
mentioned
in
\cite{deser}.

In addition to the inconvenience in technical terms, there is a more significant problem, the nonlinearity of the equations of motion in the weak gravity limit contradicts the principle of superposition for the gravitational field. Indeed, for the correction $h_{\m\n}$ to the flat metric corresponding (in the sense of Regge-Teitelboim
equation
\eqref{15a1})
to
the
sum
of
two
contributions
to
the
EMT
of
ordinary
matter
to
be
equal
to
the
sum of the corrections $h_{\m\n}$ corresponding to each of the contributions to the EMT, the factor $b^a_{\m\n}$ in
\eqref{15a1}
must be nonzero in the zeroth order of the expansion in terms of $\de y^a$. This must in fact be the case for each of
the
six
nontrivial
equations
\eqref{15a1}:
it
must
be
taken
into
account
that,
the
second
fundamental
form
$b^a_{\m\n}$
is by definition orthogonal with respect to the index $a$ to the four vectors $\dd_\m y^a$ that are tangent to the surface,
and
hence,
among
the
ten
equations
in
\eqref{15a1},
four
are
satisfied
identically
at
each
point.
As
a
result,
we conclude that superposition principle for the gravitational field requires that the second fundamental form $\bar b^a_{\m\n}$
corresponding to the background embedding function $\bar y^a$ have rank 6 if it is considered as a $10\times10$ matrix with the indices $a$ and $(\m\n)$ (we note that $b^a_{\m\n}$
is symmetric with respect to the permutation of $\m,\n$ and
this
pair
of
indices
can
be
replaced
with
a
multi-index
taking
ten
values).
For
trivial
embedding
\eqref{nnn4},
it turns out that $\bar b^a_{\m\n}=0$, and hence this condition for the rank is not satisfied.

Embeddings for which the $b^a_{\m\n}$ have the highest possible rank (for the considered dimensions of the
surface and the ambient space, it is equal to 6) can be called \emph{unfolded}, because in this case, the surface in the ambient space occupies a subspace of the maximum possible dimension: it cannot be "unfolded" any more. This corresponds to the generic case because reducing the rank of $b^a_{\m\n}$ amounts to some additional
condition and corresponds to a set of measure zero. The embeddings of flat metrics with the property of "unfoldedness"
were
studied
in
\cite{statja71};
for
spherical
symmetry,
such
an
embedding
was
proposed
in
\cite{statja76}.

If the background embedding function $\bar y^a$ is an unfolded one, then the relation between the corrections $h_{\m\n}$ to the flat metric and $\de y^a$ to the background embedding function is linear. To see this, we decompose an arbitrary quantity $\de y^a$ into the longitudinal and transverse parts with respect to the background surface:
\disn{p4}{
\delta y^a=\xi^\m\dd_\m \bar y^a+\delta y^a _{\perp}, \qquad \delta y^a _{\perp}\dd_\m \bar y_a=0.
\nom}
From
\eqref{r1},
in
the
first
order
in
$\de y^a$ we then find
\disn{nnn6}{
h_{\mu \nu} =  (\partial _\mu \bar y_a)(\partial _\nu \delta y^a)+(\partial _\m \delta y^a)(\partial _\n \bar y_a) =
\dd_\m\xi_\n+\dd_\nu\xi_\m  -2 \bar b^a _{\mu \nu} \delta y_{a \perp},
\nom}
where
the
definition
and
properties
of
the
second
fundamental
form
are
used
(see,
e.g.,
\cite{statja18}
for
more
details).

If $\bar b^a_{\m\n}$ has rank 6, then all the 6 components $\delta y^a_\perp$ can be found from this relation, which means a linear
.. . relation between all components $h_{\mu \nu}$ and $\delta y_a$. In the case of a lower rank, it is impossible to find some components $\delta y^a_\perp$ from the linearized relation, which means that they are related to $h_{\mu \nu}$ nonlinearly.

In summary, we conclude that the background embedding corresponding to the Minkowski space is to be taken in the form of an unfolded embedding the corresponding metric. The Regge-Teitelboim equations are still linearizable on such a background; in the first order, we then than have linearity in corrections to both the embedding function and the metric. The linearized equations have the form
\disn{nnn7}{
\ls G^{\m\n}_\text{lin}-\ka\, T^{\m\n} \rs \bar b^a_{\m\n}=0
\nom}
(where $G^{\m\n}_\text{lin}$ the linearized Einstein tensor, standardly expressed in terms of $h_{\mu \nu}$); they represent six (as
a result of the above-mentioned transversality $\bar b^a_{\m\n}$ with respect to the subscript $a$) linear combinations of
the ten Einstein equations.

Linearized
equations
\eqref{nnn7}
were
studied
in
\cite{statja76}
in
the
case
of
spherical
symmetry.
If we restrict ourself to the linear approximation, then the solution involves an arbitrariness corresponding to the choice of the distribution of fictitious matter. This arbitrariness can be limited if the second-order equation is taken into account; under the assumption of a static metric (which physically corresponds, for example, to the .nal stage of galaxy formation), a nonlinear equation for the linear approximation parameters then arises. As a result, we have shown that the arbitrariness in the choice of the background embedding can be chosen such that the resultant gravitational potential agrees well with the observed DM distribution in galactic halos (neglecting deviations from spherical symmetry for real galaxies).

\section{Nonrelativistic limit}
As
mentioned
in
Sec.~3,
in
the
embedding
theory,
fictitious
matter
can
be
characterized
by
the
set
of
currents $j^\m_a$,
which
are
conserved
(in
the
sense
of
Eqs.~\eqref{nnn1})
and
\eqref{r6})
as
a
consequence
of
the
equations
of

a motion. In
the
limit
studied
in
\cite{statja67,statja68},
these
vectors
become
nonrelativistic
if
the
$j^\m_a$ are regarded as four
ambient-space vectors labeled by the superscript $\m$:
 \disn{s7}{
j^\m_a=\de^0_a j^\m+\de j^\m_a,\qquad
\de j^\m_a\to0.
\nom}
If
the
metric
becomes
flat
simultaneously
with
taking
this
limit,
then
it
can
be
shown
\cite{statja68}
that
in
some
coordinates and for a finite interval of time $x^0$, the limit background embedding function has the form
 \disn{n13}{
\bar y^0=x^0,\qquad \bar y^I=\bar y^I(x^i)
\nom}
(here and hereafter, $i,k,\ldots=1,2,3$ and $I,K,\ldots=1,\ldots,9$), where $\bar y^I(x^i)$ is an unfolded embedding of the three-dimensional Euclidean metric into ten-dimensional Euclidean space. Because the three-dimensional metric of the general form has six independent components, such an embedding is parameterized by $9-6=3$ arbitrary functions.

The limit EMT of fictitious matter then becomes
 \disn{s25}{
\bar\ta^{\m\n}=\bar\rho_\ta \de^\m_0 \de^\n_0,
\nom}
that is, in the limit taken at times from a finite interval of $x^0$, this matter is at rest, which clari.es the result that follows by taking the limit in the action (see the end of Sec.~3).
The equations of motion require the density of fictitious matter to not change with time (again in a range of finite values of $x^0$), but it can depend arbitrarily on the spatial coordinates $x^i$; this dependence is determined by the choice of initial data. In the limit, fictitious matter is dust-like and at rest, and therefore it is nonrelativistic before taking the limit.

The
second
fundamental
form
corresponding
to
embedding
function
\eqref{n13}
is
given
by
\disn{nnn8}{
\bar b^a_{\m\n}=\de^a_I\de^i_\m\de^k_\n\,\bar b^I_{ik},
\nom}
where $\bar b^I_{ik}$ is transverse with respect to the superscript $I$ (i.e., $b^I_{ik}\dd_m \bar y_I=0$) and, as a consequence of the
assumed unfolded embedding, it has rank 6 if understood as a $9\times6$ matrix with indices $I$ and $(ik)$. This
allows introducing the quantity $\bar\al^{ik}_a$ that is its inverse in a certain sense; it is uniquely defined by the
relations
 \disn{s16}{
\bar\al^{ik}_I=\bar\al^{ki}_I,\qquad
\bar\al^{ik}_I \dd_m y^I=0,\qquad
\bar\al^{ik}_I \bar b_{lm}^I=\frac{1}{2}\ls \de^i_l\de^k_m+\de^i_m\de^k_l\rs.
\nom}
For a finite range of $x^0$, both this quantity and $\bar y^I$ are independent of time.

However, when taking the relativistic limit, it is necessary to take the speed of light $c$ to infinity, while
it relates $x^0$ to the nonrelativistic time $t$ by the standard formula $x^0=ct$. As a result, the finite range
of $t$ then corresponds to an unbounded range of $x^0$, and hence the quantity $\bar y^I$ acquires a dependence on the time $t$. At every time instant, it remains an embedding of the three-dimensional Euclidean metric into nine-dimensional Euclidean space, which means that it undergoes isometric bending as time progresses (we recall that such an embedding is parameterized by three functions, but it is unfortunately impossible to write such a parameterization explicitly). With respect to $x^0$, we have a nonuniform convergence of the background
embedding
function
to
its
form
in
\eqref{n13}:
it
converges
only
for
each
finite
interval
of
$x^0$. We note
that along with $\bar y^I$, $\bar\al^{ik}_I$ also acquires a dependence on $t$.

If we assume that ordinary matter is dust-like and moves slowly, then it can be described by a density $\rho$. In
the
nonrelativistic
limit,
the
equations
of
motion
of
the
embedding
theory
then
reduce
\cite{statja67,statja68}
to
the
Poisson equation for the Newtonian gravitational potential $\ff$,
 \disn{s53a4}{
\Delta\ff=4\pi G (\rho+\rho_\ta)
\nom}
(where $G$ is Newton's gravitational constant), and to the nonrelativistic equations of motion of fictitious matter,
 \disn{u9.1}{
\dd_t \ps=\ff+\frac{1}{2}\ga^I\ga^I,
\nom}\vskip -2em
 \disn{u9.2}{
\dd_t \bar y^I=\ga^I,
\nom}\vskip -2em
 \disn{u9.3}{
\dd_t \rho_\ta=-\dd_i (\rho_\ta v_\ta^i),
\nom}\vskip -2em
 \disn{u9.4}{
\dd_t (\rho_\ta v_\ta^m)=-\rho_\ta \dd_m\ff+
\dd_l \Biggl(\rho_\ta\bar\al^{lm}_I \biggl(\bar\al^{ik}_I\Bigl((\dd_i \ga^L)(\dd_k \ga^L)+\dd_i\dd_k\ff\Bigr)+2v_\ta^i\dd_i \ga^I\biggr)\Biggr),
\nom}
where $\dd_t\equiv\dd/\dd t$ and
 \disn{u8}{
\ga^I=(\dd_i \ps)\dd_i y^I+\bar{\al}^{ik}_I \dd_i\dd_k \ps.
\nom}
In this nonrelativistic limit, fictitious matter is described by its density $\rho_\ta=\ta^{00}$ and velocity $v^i_\ta=c \ta^{0i}/\rho_\ta$
and also by four variables that can be called "geometric": a scalar function $\ps$ and three implicitly present functions parameterizing the embedding $\bar y^I$ of the three-dimensional Euclidean metric into nine-dimensional Euclidean space at each instant of time. With
\eqref{u8},
it
can
be
verified
that
Eq.~\eqref{u9.2}
guarantees
that
as
time progresses, the embedding function experiences just an isometric bending. Thus, fictitious matter is described by eight variables, corresponding to four pairs of canonically conjugate variables, as suggested by
the
canonical
analysis
at
the
end
of
Sec.~2.

System
of
equations
\eqref{u9.1}-\eqref{u9.4}
is
naturally
divided
into
two
pairs.
The
first
pair,
Eqs.
\eqref{u9.1}
and
\eqref{u9.2},
determines the dynamics of four "geometric" variables and does not contain the "physical" variables $\rho_\ta$ and $v^i_\ta$. The
second
pair,
Eqs.
\eqref{u9.3}
and
\eqref{u9.4},
determines
the
dynamics
of
"physical" variables, but there
is
no
complete
separation
of
variables
because
the
"geometric"
variables
enter
\eqref{u9.4}
via
the
quantities
$\bar{\al}^{ik}_I$ and $\ga^I$. The physical meaning of
Eq.~\eqref{u9.3}
is
clear:
it is the continuity equation for fictitious matter. The
physical
meaning
of
Eq.~\eqref{u9.4}
can
be
clarified
if
this
equation
is
rewritten
in
the
form
 \disn{u11}{
\rho_\ta (\dd_t  + v_\ta^i\dd_i ) v_\ta^m=
-\rho_\ta \dd_m\ff+
\dd_l \Biggl(\rho_\ta\bigg[v_\ta^l v_\ta^m+\bar\al^{lm}_I \biggl(\bar\al^{ik}_I\Bigl((\dd_i \ga^L)(\dd_k \ga^L)+\dd_i\dd_k\ff\Bigr)+2v_\ta^i\dd_i \ga^I\biggr)\bigg]\Biggr).
\nom}
In this expression, the acceleration of an "individual particle" of fictitious matter appears in the left-hand side (we recall that there are no such particles from the fundamental standpoint, but it is convenient to discuss fictitious matter in such terms), and then the right-hand side of the equation is the force (per unit volume) acting on this particle. The first term is the usual gravitational force corresponding to the Newtonian approximation, and the remaining terms are some self-action force of fictitious matter, which depends
not only on the "physical" characteristics of this matter (the density $\rho_\ta$ and the velocity $v^i_\ta$) but also on its "geometric" characteristics $\ps$ and $y^I$. If the $1/c$ corrections are taken into account, the embedding function corresponding
to
the
nonrelativistic
approximation
has
the
form
\cite{statja67,statja68}
 \disn{s68}{
y^0=c\,t+\frac{1}{c}\,\ps\ls t,x^i\rs+o\ls\frac{1}{c^2}\rs,\no
y^I=\bar y^I\ls t,x^i\rs+\frac{1}{c^2}\bar\al^{Iik}\ls\frac{1}{2}(\dd_i \ps)(\dd_k \ps)-\ff\de_{ik}\rs+o\ls\frac{1}{c^2}\rs.
\nom}

In the embedding theory, the presence of the self-action for fictitious matter allows us to hope that in the
case
of
its
identification
with
DM,
the
"core-cusp"
problem
mentioned
in
Sec.~1
can
be
solved.
In solving this
problem,
the
method
of
analytic
evaluation
of
the
matter
distribution
profile
\cite{statja73}
can
be
used
to
avoid
time-consuming simulations taking deviations from the Newtonian behavior of fictitious matter particles due to self-action into account. The method amounts to considering the particle distribution function along possible trajectories of motion and finding the relation between the behavior of the profile and the asymptotics at zero of the particle distribution function with respect to the absolute value of the angular momentum.

\section{Conclusions}
The
modified
theory
of
gravity
proposed
by
Regge
and
Teitelboim
\cite{regge},
the
embedding
theory,
has
a simple geometric meaning, which amounts to the assumption that our space-time is a four-dimensional surface in a flat ambient space. On the other hand, the embedding theory can be understood as a result of changing
the
independent
variable
in
the
GR
action,
Eq.~\eqref{r1},
which
relates
this
theory
to
mimetic
gravity,
a
theory
popular
in
recent
years
\cite{mukhanov},
and
to
other
modifications
of
gravity
arising
as
a
result
of
changes
of variables involving derivatives. The advantage of the embedding theory over alternative versions is the clear geometric meaning of the formula for the change of variables.

In the weak gravitational field limit, the principle of superposition for gravity requires that an unfolded surface
\cite{statja71}
be
used
as
the
background
embedding
(corresponding
to
the
Minkowski
space
metric).
Finding
the explicit form of all unfolded embeddings for the Minkowski space metric remains an unsolved mathematical problem.

When reformulating the embedding theory in the GR form with fictitious matter, the emerging fictitious matter has a sufficiently large number of degrees of freedom: there are four of them, which allows arbitrarily setting eight initial functions at the initial time instant. Such a large number of degrees of freedom allows us to hope to successfully explain many effects usually associated with DM; this is in contrast to mimetic gravity, which requires further complicating the theory by introducing new terms to obtain the same result. For the embedding theory, which turns out to be sufficiently complex originally, there is a hope to do this without additional complications; but the complexity of its equations of motion makes it difficult to derive consequences that could be compared with DM observations.

On the scale of galaxies, this problem reduces to analyzing solutions of nonrelativistic equations \eqref{u9.1}-\eqref{u9.4}. It is necessary to understand the properties of the self-action force of fictitious matter and the distribution profile of fictitious matter in galaxies after the inclusion of fictitious matter. Finding the behavior of this profile from the equations in the domain of galactic halos would allow comparing the results with the observed galaxy rotation curves, and finding its behavior at the center of a galaxy would indicates whether
the
"core-cusp"
problem
mentioned
in
Sec.~1
can
be
solved
in
the
framework
of
this
approach.

It is also necessary to study the behavior of fictitious matter on cosmological scales, under the assumptions of uniformity and isotropy of the distribution of ordinary matter, which are usual for Friedmann models.
Such
a
study
was
conducted
in
\cite{davids01,statja26}.
It was shown that for some choice of the initial ratio of the amount of fictitious to ordinary matter (the latter containing the cosmological constant), it is possible to obtain
the
ratio
observed
at
present
\cite{davids01};
however,
for
a
natural
ratio
taken
at
the
beginning
of
the
inflation
epoch,
the
contribution
of
fictitious
matter
turns
out
to
be
strongly
suppressed
at
the
present
time
\cite{statja26}.

We emphasize that in both papers sited above, the embedding into a five-dimensional ambient space was chosen as a surface corresponding to the metric of the Friedmann model. If we assume that the ambient space of the embedding theory is ten-dimensional, then in the framework of the symmetry approximation of the Friedmann model, the surface is embedded into a five-dimensional subspace of the ambient space. As a consequence, the corresponding embedding is not an unfolded one. If this is true not only "on average" (on very large scales) but also on small scales, then the principle of superposition for the gravitational field is not
satisfied
(see
Sec.~4),
which
contradicts
the
observations.
Therefore,
we
have
to
assume
that
only
some
averaged surface lies in a five-dimensional subspace of the ambient space, and a more precise analysis on small scales shows the exact surface "goes out" of that subspace, i.e., deviations in the transverse directions become noticeable. Because of the nonlinearity of the Regge-Teitelboim equations, it seems unlikely that the results obtained for a precisely five-dimensional embedding would then remain true. It is therefore necessary to repeat the study presented here, either including the extra dimensions into which the surface "goes out" on small scales or originally choosing the embedding for the Friedmann metric that is also unfolded on cosmological scales.

{\bf Acknowledgements.}
The author is grateful to the organizers of the VII International Conference "Models of quantum field theory" (MQFT-2022) dedicated to the 82th anniversary of Professor Alexandr Nikolaevich Vasiliev and to the 80th anniversary of Professor Vladimir Dmitrievich Lyakhovski.


\end{document}